\documentclass[aps,prc,superscriptaddress,showpacs,nofootinbib,floatfix,twocolumn]{revtex4}

\usepackage{doi,graphicx,amssymb,amsmath,dcolumn}
\newcommand{\unit}[1]{\ensuremath{\, \mathrm{#1}}}
\newcommand{\etal}{\textit{et al. }}

\newcolumntype{d}[1]{D{.}{.}{#1} }
\begin{document}

\title{Pasta Nucleosynthesis: Molecular dynamics simulations of nuclear statistical equilibrium}
\author{M. E. Caplan}\email{mecaplan@indiana.edu}
\author{A. S. Schneider}\email{andschn@indiana.edu}
\author{C. J. Horowitz}\email{horowit@indiana.edu}
\affiliation{Department of Physics and Nuclear Theory Center, Indiana University, Bloomington, IN 47405, USA}
\author{D. K. Berry}\email{dkberry@iu.edu}
\affiliation{University Information Technology Services, Indiana University, Bloomington, IN 47408, USA}
\date{\today}
\begin{abstract}

\begin{description}
\item[Background]  Exotic non-spherical nuclear pasta shapes are expected in nuclear matter at just below saturation density because of competition between short range nuclear attraction and long range Coulomb repulsion.
\item[Purpose]
We explore the impact of nuclear pasta on nucleosynthesis, during neutron star mergers, as cold dense nuclear matter is ejected and decompressed. 
\item[Methods]
 We perform classical molecular dynamics simulations with 51\,200 and 409\,600 nucleons, that are run on GPUs.  We expand our simulation region to decompress systems from an initial density of $0.080\unit{fm}^{-3}$ down to $0.00125\unit{fm}^{-3}$. We study proton fractions of $Y_{P}=$ 0.05, 0.10, 0.20, 0.30, and 0.40  at $T =$ 0.5, 0.75, and 1.0 MeV. We calculate the composition of the resulting systems using a cluster algorithm.
\item[Results]
 We find final compositions that are in good agreement with nuclear statistical equilibrium models for temperatures of $0.75$ and $1 \unit{MeV}$.  However, for proton fractions greater than $Y_{P}=0.2$ at a temperature of $T = 0.5 \unit{MeV}$, the MD simulations produce non-equilibrium results with large rod-like nuclei. 
 \item[Conclusions]
Our MD model is valid at higher densities than simple nuclear statistical equilibrium models and may help determine the initial temperatures and proton fractions of matter ejected in mergers.
\end{description}


\end{abstract}



\maketitle

\section{Introduction}\label{sec:Intro}

Neutron-rich matter forms during a core-collapse supernova, as an increasing electron Fermi energy drives electron capture. Furthermore, while the outer crust of a neutron star consists of an ion lattice, the core (or at least the outer core) is believed to be uniform nuclear matter.  Between these two phases, a transition layer likely exists that involves non-spherical shapes \cite{Pethick19987, PhysRevLett.50.2066}. 

Near nuclear saturation density, the system is \textit{frustrated} because of an inability to minimize all of the fundamental interactions. This competition between the attractive short-range nuclear force, with a range of order  $1\unit{fm}$, and the repulsive long-range Coulomb force produces complex nonuniform structures, called \textit{nuclear pasta} \cite{PhysRevC.69.045804}. Theoretical symmetry arguments and numerical simulations of the phases of nuclear pasta have identified a variety of structures. These phases are, in order of increasing density: spheroids (``gnocchi''), rods (``spaghetti''), slabs (``lasagna"), uniform matter with cylindrical voids (``anti-spaghetti''), and uniform matter with spherical voids (``anti-gnocchi'') \cite{Okamoto2012284, PhysRevC.88.065807}. These geometries have been produced both by large-scale classical molecular dynamics simulations and quantum Hartree-Fock calculations \cite{Schuetrumpf:2014aea, PhysRevC.79.055801}. 

More exotic structures have also been recently identified, such as flat plates with a lattice of holes or ``nuclear waffles'' \cite{2014arXiv1409.2551S}, a networked ``gyroid'' phase \cite{Schuetrumpf:2014aea, PhysRevLett.103.132501} and chiral deformations in intertwined lasagna configurations \cite{Alcain:2014fma}.  This indicates that pasta can be described by not only the simple shapes presented above, but that for varying proton fractions, temperatures, and densities there appear to be a richer variety of possible structures.  Increases in computing power now facilitate the study of a wider range of pasta shapes. 

As nuclear pasta is expected to form during the core collapse phase of a supernova, it may play an important role in the evolution of the supernova and the resulting neutron star \cite{Horowitz:2012za}. 
For example, supernova neutrinos can scatter coherently from pasta, because neutrino wavelengths are comparable to the pasta size. Therefore, calculations of the static structure factor of nuclear pasta can help determine the neutrino opacity in core collapse supernovae \cite{PhysRevC.69.045804, Alcain:2014fma}. 

Furthermore, the pasta shapes could influence the thermal and electrical conductivities of the inner crust of neutron stars.  Horowitz \etal suggest that topological defects in nuclear pasta increase electron pasta scattering and reduce the thermal conductivity.  This could slow crust cooling after accretion in low mass X-ray binaries  \cite{2014arXiv1410.2197H}.    Furthermore defects would also reduce the electrical conductivity.  This could lead to the decay of magnetic fields, if the fields are supported by currents in the crust.   As a result, neutron stars would stop spinning down \cite{2013NatPh...9..431P}. 


The elastic properties of pasta, such as its shear modulus and breaking strain, help determine the maximum size of  ``mountains'' that may be present on neutron stars.  Here pasta may be a particularly important part of the crust because of its high density.  These mountains, on rapidly rotating stars, are energetic sources of gravitational wave radiation \cite{PhysRevLett.102.191102}.  Finally the pasta breaking strain is relevant for crust breaking models of star quakes and magnetar giant flares \cite{CTPP:CTPP201100075}.


However, despite the large literature on nuclear pasta and its relevant astrophysical properties, very little work has been done identifying the role pasta might play in heavy element nucleosynthesis.  Astrophysical sites may involve the ejection of dense neutron rich matter that may originally be in a nuclear pasta phase.  As the matter decompresses, the pasta shapes may react to form seed nuclei and free neutrons, and these may later undergo more conventional nucleosynthesis reactions.   

It is thought the rapid neutron capture process, or r-process, produces about half of the elements heavier than iron, but the site of the r-process remains uncertain \cite{2007PhR...450...97A}. Previously, supernovae were prime candidates but the most recent simulations of the neutrino driven wind in core collapse supernova are not neutron rich enough to produce heavy r-process elements \cite{PhysRevLett.82.5198,PhysRevD.65.083005,2007PhR...450...97A}.

Neutron star mergers have recently been identified as strong candidate sites for the r-process due to their ejection of neutron rich matter and their relatively high galactic merger rate, which is now expected to be as high as $\sim \,$ $10^{-4} \unit{yr}^{-1}$ \cite{2007PhR...450...97A,lrr-2008-8}.
Recently, NS-NS mergers and NS-BH mergers have been studied using relativistic hydrodynamic simulations, which find that the nuclear abundances in ejecta match well to solar ratios and are robust for a variety of mass ratios for the merging system \cite{2011ApJ...738L..32G, 2012MNRAS.426.1940K}.  Calculations show that the inner crust provides the largest portion of ejecta mass and that the amount of ejecta varies with the mass ratio of binary.  Estimates range between $10^{-3} \unit{M_{\odot}}$ for symmetric NS mergers and $10^{-2} \unit{M_{\odot}}$ for asymmetric NS mergers, with systems where $M_{1}/M_{2} = 0.55 $ ejecting greater than $0.2 \unit{M_{\odot}}$ \cite{MNR:MNR10238}. 

In this paper, we explore the possibility that nuclear pasta, ejected from the inner crust during these mergers, could provide the initial material for the r-process.  Pasta properties could be important for the evolution of the material's temperature and entropy.  Furthermore, weak interactions in the pasta, such as neutrino or charged lepton capture, will determine the evolution of the proton fraction.



In previous work we identified the average size of gnocchi (mass number of nuclei) using MD simulations \cite{PhysRevC.88.065807}.  In that work we evolved dense matter with a proton fraction of $Y_P=0.40$ at a temperature of 1\unit{MeV} from high to low densities, $n=0.10\unit{fm}^{-3}$ to $n\sim0.01\unit{fm}^{-3}$, by expanding the simulation volume at different rates. That work observed the nucleation mechanism for a number of different pasta phase transitions and quantified those transitions by calculating the average mean and Gaussian curvatures, allowing us to characterize the phases by Minkowski functionals. 

The simulated NS-NS mergers cited above observed ejecta expansion timescales on the order of milliseconds \cite{MNR:MNR10238}, which is much slower than the nuclear timescales in this work. As we expect nuclear matter to remain in nuclear statistical equilibrium (NSE) while evolving over such long timescales, we expand our simulation volume as slow as possible for as many time steps as computationally allowable, and compare to faster expansion rates to confirm that we are expanding slow enough to remain in quasi-static equilibrium, which should be expected of ejecta with millisecond expansion timescales. 

There have been many MD simulations of heavy ion collisions to study the formation of clusters via multi-fragmentation, see for example \cite{PRC63.024602,PLB519.46}.  Typically these involve relatively small systems and sometimes neglect Coulomb interactions.  In this paper we use the recently developed Indiana University Molecular Dynamics GPU code (IUMD) to simulate larger volumes of nuclear matter as its density decreases.  We include full Coulomb interactions and consider a range of proton fractions and temperatures.   

In Sec. \ref{sec:Formalism} we describe our MD formalism and discuss the IUMD GPU code and its performance.   In Sec. \ref{sec:Results} we discuss our results and compare to previous work. 
We conclude in Sec. \ref{sec:Conclusions}.

\section{Formalism}\label{sec:Formalism}

To describe the decompression of ejected matter during a neutron star merger we perform molecular dynamics (MD) simulations with a fixed number of nucleons in a volume that slowly expands.   This corresponds to decreasing densities from 0.08 fm$^{-3}$, where complex nuclear pasta phases may be present, to 0.01 fm$^{-3}$ or below, where many nucleons bind into largely isolated nuclei.  We first discuss our MD formalism in Sec. \ref{ssec:Formalism}, 
and then describe the IUMD GPU code and its performance in Sec. \ref{ssec:codes}.  Finally we review a cluster algorithm in Sec. \ref{ssec:clusteralgorithm} that determines which nucleons belong to which nuclei.   This algorithm allows us to deduce the final nuclear abundances produced by our simulations.   

\subsection{Semiclassical nuclear pasta model}\label{ssec:Formalism}

Our MD formalism is the same as that used by Horowitz \etal in previous works 
\cite{PhysRevC.69.045804,PhysRevC.70.065806,PhysRevC.72.035801,PhysRevC.78.035806,PhysRevC.86.055805,GiménezMolinelli201431,PhysRevC.88.065807,PhysRevC.89.055801} and is briefly reviewed here. In particular, much of this work is a continuation of Schneider \etal \cite{PhysRevC.88.065807}.

Our simulation volume is a cubic box with periodic boundary conditions which contains point-like protons and neutrons, with mass $M=939\unit{MeV}$.  The electrons are assumed to form a degenerate relativistic Fermi gas and are not explicitly included in the simulations. The protons and neutrons interact via the two-body potentials:

\begin{subequations}
\begin{align}
 V_{np}(r)&=a e^{-r^2/\Lambda}+[b-c]e^{-r^2/2\Lambda}\\
 V_{nn}(r)&=a e^{-r^2/\Lambda}+[b+c]e^{-r^2/2\Lambda}\\
 V_{pp}(r)&=a e^{-r^2/\Lambda}+[b+c]e^{-r^2/2\Lambda}+\frac{\alpha}{r}e^{-r/\lambda}.
\end{align}
\end{subequations}

The indices indicate whether the potential describes a neutron-proton interaction, a neutron-neutron interaction, or a proton-proton interaction, while $r$ is the separation between the two nucleons. The constants $a$, $b$, $c$, and $\Lambda$ can be found in Table \ref{Tab:parameters} and are chosen to approximately reproduce some bulk properties of pure neutron matter and symmetric nuclear matter, as well as the binding energies of selected nuclei \cite{PhysRevC.69.045804}.

\begin{table}[h]
\caption{\label{Tab:parameters} Nuclear interaction parameters. 
The parameter $a$ defines the strength of the short-range repulsion between nucleons, 
$b$ and $c$ the strength of their intermediate-range attraction and $\Lambda$ the length scale of the nuclear potential.}
\begin{ruledtabular}
\begin{tabular}{*{4}{c}}
$a$ (MeV) &$b$ (MeV)&$c$ (MeV)&$\Lambda$ (fm$^{2}$) \\
\hline
  110     &  $-$26    &   24    &    1.25      \\
\end{tabular}
\end{ruledtabular}
\end{table}

The proton-proton interaction includes the Coulomb repulsion that is screened by the electron gas.  This screening has a characteristic length $\lambda$ that depends on the fine structure constant $\alpha$ and the electron Fermi momentum $k_F=(3\pi^2n_e)^{1/3}$, where $n_e$ is the electron density (assumed equal to the proton density) and the electron mass is $m_e$.  Its value is
\begin{equation}
 \lambda=\frac{\pi^{1/2}}{2\alpha^{1/2}}\left(k_F\sqrt{k_F^2+m_e^2}\right)^{-1/2}.
\label{eq:lambda}
\end{equation}
Though in this work we fix $\lambda$ to the slightly smaller value  $10\unit{fm}$, for all proton fractions, to agree with the value used in earlier work.  We do not expect our results to be very sensitive to the precise value of $\lambda$.

We use a cut-off radius for the nuclear potential of 11.5 fm, and no cutoff radius for the Coulomb potential. 
The nuclear potential is assumed to be zero and is not computed for separations greater than the cut-off distance. The boundary conditions allow particles to only interact with the nearest periodic image of other nucleons. Due to the short range of the nuclear potential, the computation of the nuclear force can be greatly accelerated with the periodic construction of neighbor-lists, which are discussed in detail in Sec. \ref{sssec:IUMD}. 

After computing all the inter-particle forces, the nucleon positions and velocities are updated using a velocity-Verlet algorithm \cite{PhysRev.159.98}. After the timestep $\Delta t$ is completed, the box size is increased by a small amount. The length of each side of the box $l_{i}$ ($i = x, y, z$) and the volume $V$ is
\begin{subequations}
\begin{align}
l_{i}(t) = l_{i}(0) \left( 1 + \dot\xi_{i} t \right) \\
V(t) = V(0) \left( 1 + \dot\xi t \right)^{3}
\label{eq:BoxLength}
\end{align}
\end{subequations}
where $l_{i}(0)$ and $V(0)$ are the initial side length and volume of the box, and $\dot\xi_{i}$ is the expansion rate. To preserve the cubic geometry, $\dot\xi_{i}$ is the same for all $i$. Particle positions and velocities are not incremented with the change in box volume, and are allowed to respond dynamically to the changing simulation volume. Furthermore, particles that cross one side of the box and reenter on the other do not have their velocities rescaled. The error this introduces is discussed in detail in Scheinder \etal \cite{PhysRevC.88.065807} and is found to be negligible for the parameters of our system because the expansion rate is very slow compared to the particle velocities. 

In order to approximately maintain an isothermal expansion, the velocities of all nucleons are rescaled every 100 time steps so that the average kinetic energy per particle is $(3/2)kT$.  

\subsection{GPU code}\label{ssec:codes}

The simulations in this work were computed using a new version of the Indiana
University Molecular Dynamics (IUMD) Fortran code, which has been modified
from our previous work Ref. \cite{PhysRevC.88.065807} to run on the GPU nodes of the Big Red II
supercomputer at Indiana University.

\subsubsection{The IUMD code}\label{sssec:IUMD}

The IUMD code has been used for a decade and run on the original Big Red
supercomputer at Indiana University (an IBM JS21), and the Kraken supercomputer
at Oak Ridge National Lab (a Cray XT5). Both these machines consisted of
general purpose multi-core CPU nodes.

In mid 2013, Indiana University acquired a Cray XE6/XK6 supercomputer. The
XE6 part of the machine consists of 344 general purpose dual 16-core CPU nodes.
The XK6 part consists of 676 accelerated nodes, containing one 16-core CPU and
one Nvidia Kepler K20 GPU \cite{BigRed}.  A new version of IUMD (version 6.3.1)
was created to take advantage of the powerful accelerated nodes. IUMD 6.3.1 is
explained in greater detail in \cite{2014arXiv1409.2551S}. Here we explain it only
enough to understand the performance we observed in our expansion runs.

IUMD is a parallel Fortran code which uses MPI to pass data between nodes,
OpenMP on each CPU to take advantage of its 16 cores, and CUDA Fortran to
take advantage of the GPUs. In discussing the code, it is helpful to think of
the two-particle interactions as making up a \emph{force matrix}, whose $ij$
element is the force $\boldsymbol{f}_{ij}$ that \emph{source} $j$ exerts on
\emph{target} $i$.  Overall of course, targets and sources are the same
particles.  In IUMD the force matrix is partitioned into a $P \times Q$ block
matrix, where $PQ$ is the total number of MPI processes (nodes). Each MPI
process is responsible for calculating forces represented by one block.  Note
that this decomposition is an abstract one, rather than one based on geometry.
Targets and sources are distributed randomly among MPI processes so that each
process is responsible for the entire simulation volume, but only a fraction
of the targets and sources in the volume. This is different from parallel
algorithms where each process is responsible for all particles in a subvolume.
The advantage of IUMD is particles do not have to be transferred from process
to process as they move from one subvolume to another.  Orchestrating such
transfers involves a level of coding complexity that may be difficult to
optimize.  The disadvantage of IUMD is that an MPI\_Allreduce must be performed to
combine partial forces from different processes to get the total force on a
target.  However, this is a single call to an MPI subroutine that hopefully
has been optimized in the MPI library. A similar MPI\_Allreduce is used to
update source particle positions each time step. Again, see \cite{2014arXiv1409.2551S}
for details.

The force calculation is by far the most time consuming part of each time step.
It consists of a short-range two-nucleon nuclear force, and a screened Coulomb
force between protons.  IUMD calculates the nuclear force on the CPUs using
an efficient neighbor list algorithm.  Each target's neighbor list consists
of those sources within distance $r_{nuc} + \delta r_{nuc}$.  Forces are
calculated only for those sources within interaction range $r_{nuc}$; however,
sources within a halo of thickness $\delta r_{nuc}$ are included in the list
so it does not have to be rebuilt every time step.  Rather, a target's list
needs to be rebuilt only when the distance it has moved, plus the maximum
distance any source on its node has moved exceeds $\delta r_{nuc}$.  For only
then is it possible that a source not in the target's list could have come
into interaction range.  Since rebuilding neighbor lists is time-consuming
and disruptive, all lists on all processes are rebuilt if any one list needs
to be.  We set $r_{nuc} = 11.5\unit{fm}$ for the runs in this paper.
Due to the rapid decrease of the nuclear force, it will rarely register
for separations greater than this, even in 64-bit IEEE arithmetic.

While the CPUs calculate the nuclear force, the GPUs calculate the screened
Coulomb force using a simple all-pairs, or \emph{particle-particle} (PP)
algorithm.  The work for each process to calculate the screened Coulomb force
scales like $(Y_P N)^2/PQ$, where $N/P$ and $N/Q$ are the number of targets
and sources on each process.  With work rising as the square of the number of
particles, the PP algorithm would seem to be very inefficient.  However, each
Kepler K20 GPU on BigRed II contains 2496 single precision floating point
cores, 832 double precision cores, and 416 special function units for
computing square root, exponential and trigonometric functions. This is a
powerful computational capacity for implementing simple, straightforward
algorithms, whereas complex algorithms may not work so well.  Thus we keep
the algorithm as simple as possible, eschewing even the use of Newton's Third
Law, as it entails branching that would slow the GPU. We depend instead on
the Kepler K20's massive parallelism.  In the next section we show that for
51 200 nucleons, $Y_p<=0.400$ and densities of interest in nuclear pasta
research, the PP algorithm running on GPUs still finishes before the neighbor
list algorithm on the CPUs.  Even for 409 600 nucleons it outperforms the
CPU at low proton fractions.

\subsubsection{Performance}\label{sssec:performance}

The PP algorithm for the Coulomb interaction takes order $O((Y_p N)^2/PQ)$
amount of work per process.  For a fixed number of nucleons, it therefore
scales as the square of the proton fraction. On the other hand, the work
involved in the neighbor list algorithm for the nuclear force is independent
of $Y_p$, but depends linearly on density, because the the number of sources
within interaction range of each target depends linearly on density.  We
therefore expect the work to calculate the nuclear force for all $N/P$ targets
on a process to be of order $O((N/P) (4\pi/3) (r_{nuc}+\delta r_{nuc})^3 (n/Q))$.
To this should be added the work to build neighbor lists.  How frequently
neighbor lists need to be rebuilt depends on the density and temperature of
the system. We chose $\delta r_{nuc} = 4.0\unit{fm}$, to reduce the frequency,
while still keeping lists relatively short.  With this value of $\delta r_{nuc}$
we found that neighbor lists were rebuilt approximately every ten timesteps for
high temperatures and densities, and every hundred time steps for low density,
low temperature, equilibrium systems. The algorithm IUMD uses to build neighbor
lists involves dividing the simulation volume into cubical cells of width
$r_{nuc}+\delta r_{nuc}$ and checking only sources in a target's cell and its
26 neighboring cells. The work in this algorithm is thus of order
$O(27(N/P)(r_{nuc}+\delta r_{nuc})^3 (n/Q))$, the same order as the nuclear
force calculation itself.  The exact coefficients in this estimate and the
nuclear force calculation were not determined, but the work to decide if a
source should go in a neighbor list is much less than calculating the force
it exerts. Thus the neighbor list build may be noticeable, but still a small
fraction of the force calculation.

We collected timing data from the simulations described in Sec. \ref{ssec:populations} and
Sec. \ref{ssec:409600} to check the above workload estimates.  These simulations involved
51 200 and 409 600 particles, run on 32 and 128 GPU nodes respectively. The
simulations were evolved for $3 \times 10^7\unit{fm/c}$ with 2 fm/c per timestep
on Big Red II, and were expanded at a rate of $\dot{\xi} = 10^{-7}\unit{c/fm}$,
starting at an initial nucleon density of $0.08 \unit{fm}^{-3}$ and decreasing
to $0.00125 \unit{fm}^{-3}$. Since the density was monotonically decreasing the
average time per MD timestep could be calculated from timestamps on checkpoint
files output by the program every $5 \times 10^5$ timesteps during the
$15 \times 10^6$ timesteps of the run.

\begin{figure}[h!]
\centering
\includegraphics[width=0.5\textwidth]{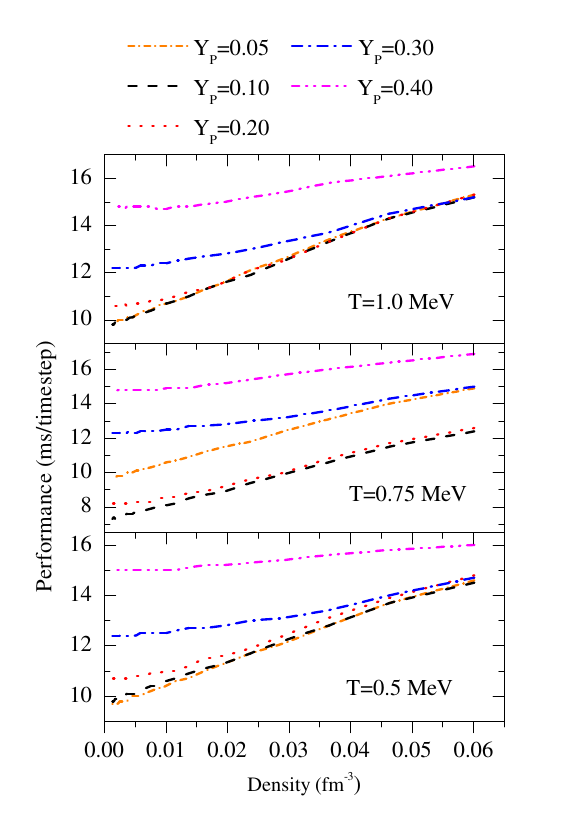}
\caption{\label{fig:performance} (Color online) Performance of the IUMD GPU code for the 15 simulations described in Secs. \ref{ssec:populations} \& \ref{ssec:nonequilibrium}. All computations were performed using 51\,200 particles with 32 GPU nodes on Big Red II.}

\includegraphics[width=0.47\textwidth]{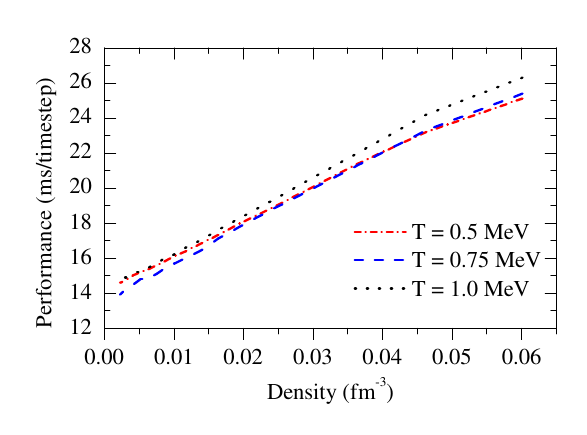}
\caption{\label{fig:performance409600} (Color online) Performance of the IUMD GPU code for the 3 simulations described in Sec. \ref{ssec:409600}. All simulations were performed using 409\,600 particles, $Y_P = 0.05$, with 128 GPU nodes on Big Red II.}

\end{figure}

Figure \ref{fig:performance} shows the average time per MD timestep as a function of density for the
five proton fractions we ran with 51 200 nucleons.  For $Y_p = 0.05$ and 0.10
time decreases almost linearly as density decreases, in accord with our model
for the work involved in the nuclear calculation. This indicates the nuclear
calculation takes longer than the Coulomb at all densities.  We call this the
linear case. For $Y_p=0.40$ time decreases linearly from $n=0.06$ to 0.01 $\unit{fm}^{-3}$,
then flattens out. For this proton fraction the nuclear force calculation takes
longer above 0.01 $\unit{fm}^{-3}$, while the Coulomb calculation takes longer
below 0.01 $\unit{fm}^{-3}$. We call this the broken linear case.

For $Y_p=0.20$ and 0.30 the decrease is linear for high density, but slower and
nonlinear at lower density.  According to our simple model, if the performance
model is not linear or broken linear, then the Coulomb calculation should take
longer at all densities, which we call the flat performance case. However, these
two proton fractions fit none of these cases.  What the model does not take into
account, is that the size of neighbor lists actually depends on local density
rather than mean density.  As nucleons cluster into pasta shapes, or nuclei at
very low mean density, the local density inside clusters goes to saturation
density, while outside is a low density gas of neutrons, a few free protons and
alpha particles.  Thus the curves for $Y_p=0.05$ and $0.10$ are linear because
there are few clusters.  For $Y_p=0.40$ clusters have not yet formed by the
time mean and local density diverge.  However, for $Y_p=0.20$ and 0.30 clusters
do form while the nuclear force calculation still dominates, so the curves
gradually flatten. Note that at about $n=0.005 \unit{fm}^{-3}$ the curve
appears to suddenly go horizontal. This may be the point at which the Coulomb
calculation finally dominates the time.

The slopes of the curves do appear to depend on proton fraction and temperature.
At $T=1.00$ MeV, neighbor lists are rebuilt more often. Since the work to do
that depends linearly on density, they increase the coefficient in the order
estimate, thus increasing the slope. 

One odd characteristic of Figure \ref{fig:performance} is that the curves do not all start at the
same point at $n=0.06 \unit{fm}^{-3}$. Other tests we did to investigate this
indicate it is due to differences in the way MPI communication is set up
between different runs. Either the node set allocated to the run, or the way
the MPI runtime system places processes on nodes, may affect the efficiency
of message passing.

In Figure \ref{fig:performance409600} we show the time per MD timestep for the $N=409 600$ nucleon run
for $Y_p=0.05$ on 128 nodes. Because of the low proton fraction, the nuclear
calculation dominates at all densities, resulting in the nearly linear decrease
in time as density decreases. Again there is a slight temperature dependence,
being greater for $T=1.00$ MeV. Finally, we note the dependence on $N$. This
run involved eight times as many nucleions on four times as many nodes (128
vs. 32). According to our model for the nuclear force, the run should have
therefore taken twice as long. Figure \ref{fig:performance409600} shows it performed better than this,
taking on average only about 1.6 times as long. This may be due to better
performance of the MPI communication for larger messages, and better performance
in the velocity Verlet update with a larger number of particles.

In general, we observe that lower proton fractions can be computed the fastest and speed up with decreasing density, while higher proton fractions take longer to compute and only slightly accelerate for decreasing densities. We observe negligible dependence on temperature. While the 409\,600 particle simulations mark an eightfold increase in the number of particles from the 51\,200 particle runs, they were only run on four times the GPU nodes. In good agreement with our prediction of performance scaling linearly with $N$, we observe the time-per-timestep to double at high density, but to only increase by $\sim 50\%$ for very low density. 



\begin{table*}[!t]
\caption{\label{Tab:expansionstats} Mass fraction of free neutrons, mean mass number of clusters ($A \geq 12$), and mean charge number of clusters from simulations of 51\,200 nucleons, expanded at four proton fractions ($Y_P$) and four stretch rates ($\dot{\xi}$) are shown. All data shown is for the final configuration of the simulation, when $n = 0.01\unit{fm}^{-3}$.
The data in the rightmost columns are offered for comparison to NSE tables. This data is generated from a Virial expansion of 8980 species of nuclei with $A\geq12$, and shows fair agreement with the IUMD results \cite{PhysRevC.82.045802}. }
\begin{ruledtabular}
\begin{tabular}{lccccccc}
\multicolumn{2}{c}{ } & \multicolumn{3}{c}{IUMD} & \multicolumn{3}{c}{NSE} \\
$Y_P$ & \multicolumn{1}{c}{ $log_{10}\dot{\xi}$} & $M_{\text{free neutrons}}$ & $\overline{A} \pm \sigma_{A}$ & $\overline{Z} \pm \sigma_{Z}$ & $M_{\text{free neutrons}}$ & $\overline{A}$ & $\overline{Z}$\\ 
\hline
0.1 & -5 & 0.6538 & 90.12 $\pm$ 24.20 & 26.25 $\pm$ 7.69 & \multicolumn{3}{c}{Unavailable} \\
 & -6 & 0.6561 & 94.08 $\pm$ 18.01 & 27.57 $\pm$ 5.69 &  & &\\
 & -7 & 0.6578 & 99.18 $\pm$ 17.42 & 29.20 $\pm$ 5.57 &  & &\\
0.2 & -5 & 0.3748 & 147.75 $\pm$ 33.88 & 47.31 $\pm$ 10.95 & 0.3335 & 179.3 & 53.80\\
 & -6 & 0.3731 & 146.22 $\pm$ 24.96 & 46.69 $\pm$ 8.30 & & &\\
 & -7 & 0.3704 & 136.12 $\pm$ 20.32 & 43.29 $\pm$ 6.72& & &\\
0.3 & -5 & 0.1359 & 186.47 $\pm$ 73.19 & 64.75 $\pm$ 24.59 & 0.0475 & 184.4 & 58.07\\
 & -6 & 0.1377 & 175.13 $\pm$ 34.48 & 60.94 $\pm$ 12.16 & & &\\
 & -7 & 0.1367 & 166.74 $\pm$ 34.71 & 57.95 $\pm$ 12.44 & & &\\
0.4 & -5 & 0.0109 & 369.37 $\pm$ 426.03 & 149.32 $\pm$ 167.45 & 0.0001 & 194.4 & 77.75\\
 & -6 & 0.0121 & 179.40 $\pm$ 29.83 & 72.64 $\pm$ 11.93 & & &\\
 & -7 & 0.0115 & 190.25 $\pm$ 29.34 & 76.94 $\pm$ 11.83 & & &\\
 & -8 & 0.0111 & 191.74 $\pm$ 24.55 & 77.55 $\pm$ 9.79 & & &
\end{tabular}
\end{ruledtabular}
\end{table*}

\subsection{Cluster Algorithm}\label{ssec:clusteralgorithm}

To find clusters of protons and neutrons in our simulations we use the Minimum Spanning Tree (MST) algorithm which we previously used in Schneider \etal \cite{PhysRevC.88.065807} and is common in other molecular dynamics studies of nuclear pasta \cite{Dorso1995197, PhysRevC.86.055805}.

The MST algorithm identifies nearest neighbors to build lists of nucleons in each cluster. The algorithm examines every pair of protons $i$ and $j$ and proton $i$ is determined to belong to a cluster $C$ if $i$ is within the cut-off distance $r_{pp}$ of at least one proton $j$ that is also a part of $C$. 
Schneider \etal found that $r_{pp} = 4.5 \unit{fm}$ was an acceptable value for all densities by examining the proton-proton correlation function $g_{pp}(r)$ \cite{PhysRevC.88.065807}. If no protons $j$ are within range of a proton $i$ then $i$ is considered its own cluster. After a list of protons in each cluster have been assembled, the neutrons in each cluster are counted. Neutrons are counted as members of a cluster $C$ if it is within a distance $r_{np}$ of at least one proton $j$ that also belongs to $C$. 
Again following Schneider \etal  we use $r_{np} = 3 \unit{fm}$. Neutrons that are not apart of any cluster are counted as free neutrons. This algorithm accounts for periodicity in the system, and checks the nearest periodic images of all protons $j$ during both the proton-proton check, and the proton-neutron check.

As neutrons are constantly being exchanged between the clusters and the free neutron gas, there exists a small probability that neutrons get miscounted if they are in the process of escaping or making close fly-bys. This probability is small, and has little effect on our results. More cluster algorithms exist, such as those that check the energy between particle pairs to check if they are bound, such as the Minimum  Spanning Tree in Two-particle Energy Space (MSTE). While they have the advantage of discriminating against questionable surface neutrons, they compare well to the MSE overall, and we find the MSE is sufficient in this work \cite{PhysRevC.88.065807, PhysRevC.86.055805, Dorso1995197}.

\section{Results}\label{sec:Results}


We start in Sec. \ref{ssec:epsdot} with a discussion of the effect of expansion rate on the final configuration of the simulation. In Sec. \ref{ssec:populations} we discuss the final populations of nuclides present in our simulations at 0.75 and 1.0 MeV. In Sec. \ref{ssec:nonequilibrium} we discuss the results of simulations at 0.5 MeV, specifically the non-equilibium effects observed at higher proton fractions. 

\begin{table*}[!t]
\caption{\label{tab:Visualizations} (Color online) Comparisons of configurations at several densities obtained from three different simulations, shown to scale. The figures are generated in Paraview by finding isosurfaces of charge density. The dark surfaces are generated where $n_{Z} = 0.03 \unit{fm}^{-3}$, and the lighter surfaces at the boundary show where $n_{Z} > 0.03 \unit{fm}^{-3}$.
The first column shows the density of the configurations in each row. }
\begin{ruledtabular}
\begin{tabular}{cccc} 
 & \bf{$T=$ 1.0 MeV} & \bf{$T=$ 1.0 MeV}  & \bf{$T=$ 0.5 MeV} \\
\bf{$n \unit{(fm}^{3}$)} & \bf{$Y_P=$ 0.1} & \bf{$Y_P=$ 0.4} & \bf{$Y_P=$ 0.4} \\
\hline
 & & & \\
 \bf{0.0601} 
& \includegraphics[width=0.2\textwidth]{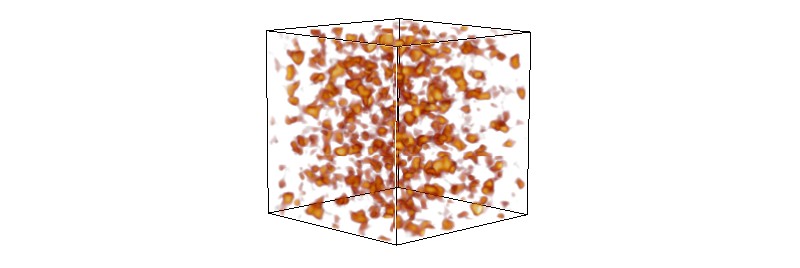}
& \includegraphics[width=0.2\textwidth]{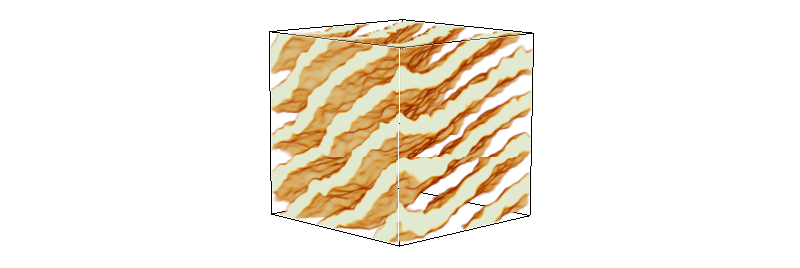} 
& \includegraphics[width=0.2\textwidth]{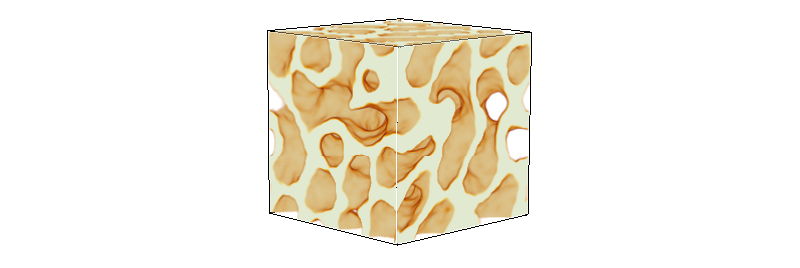}    \\
 \bf{0.0195}
& \includegraphics[width=0.2\textwidth]{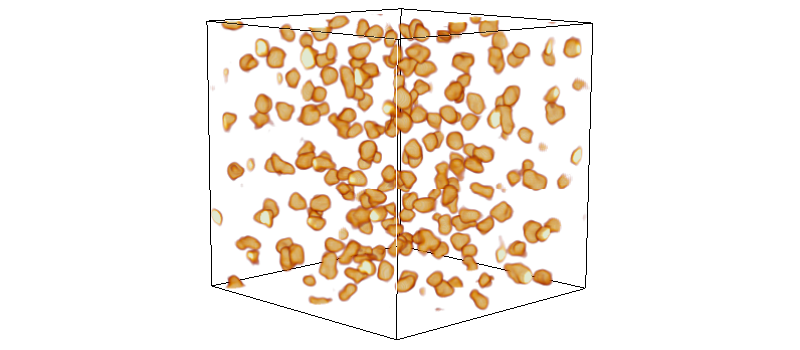}
& \includegraphics[width=0.2\textwidth]{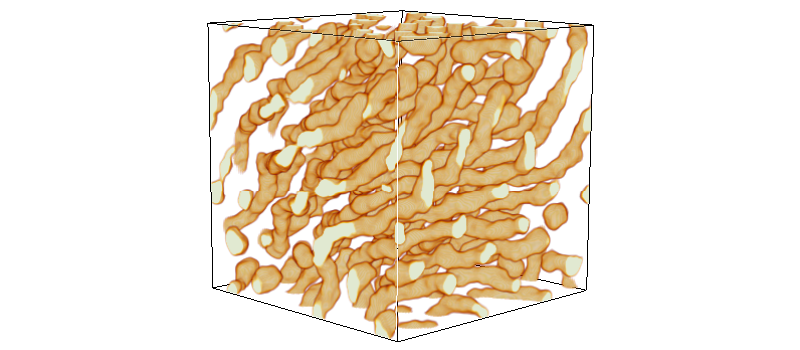} 
& \includegraphics[width=0.2\textwidth]{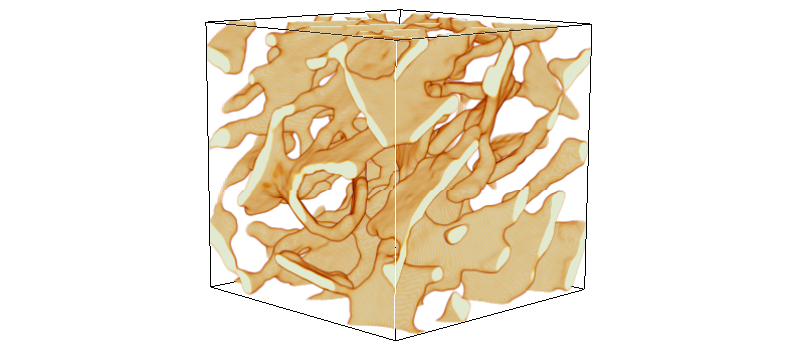}    \\
 \bf{0.0116}
& \includegraphics[width=0.2\textwidth]{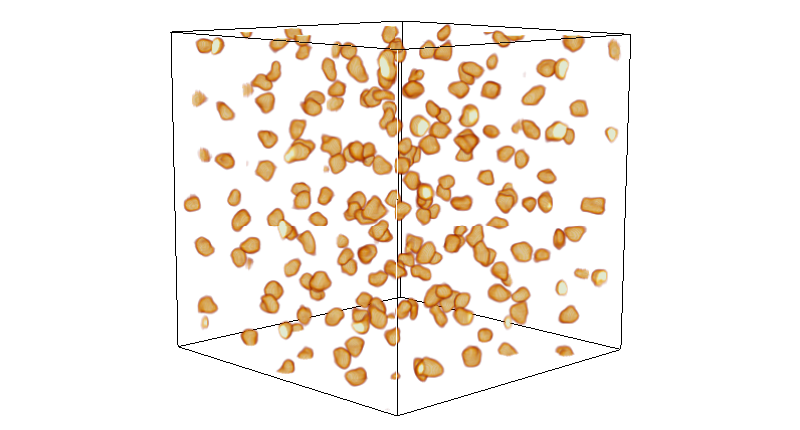}
& \includegraphics[width=0.2\textwidth]{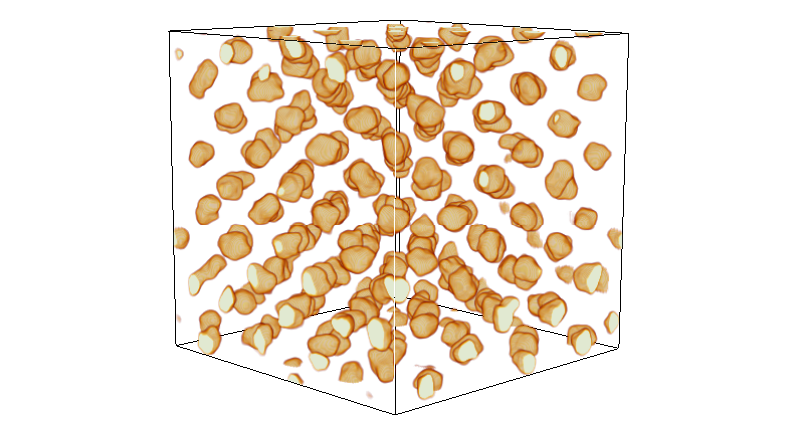} 
& \includegraphics[width=0.2\textwidth]{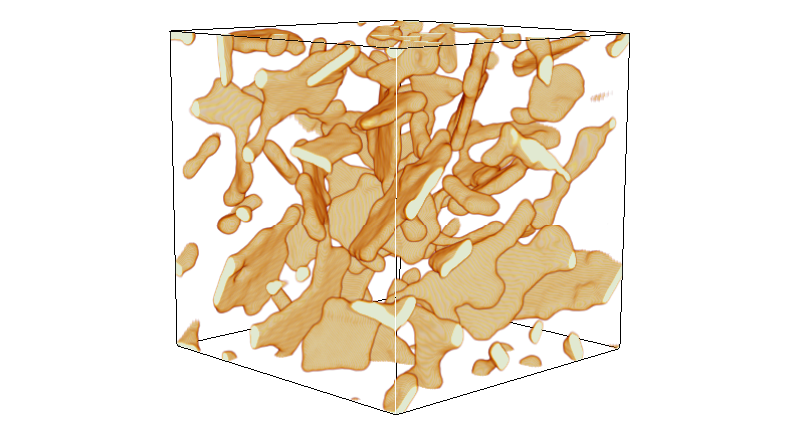}    \\
\bf{0.0027} 
& \includegraphics[width=0.2\textwidth]{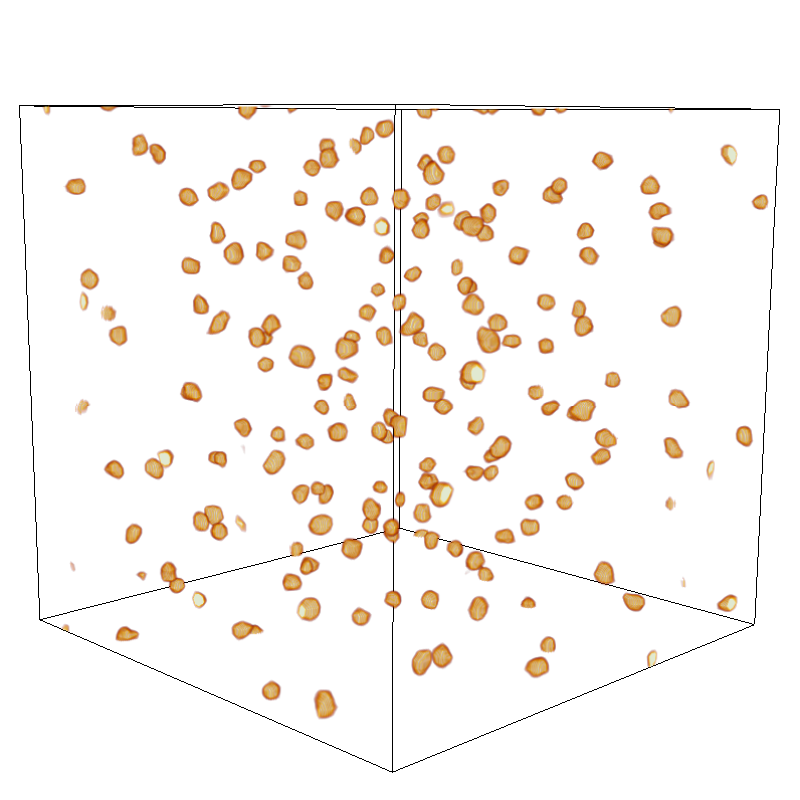}
& \includegraphics[width=0.2\textwidth]{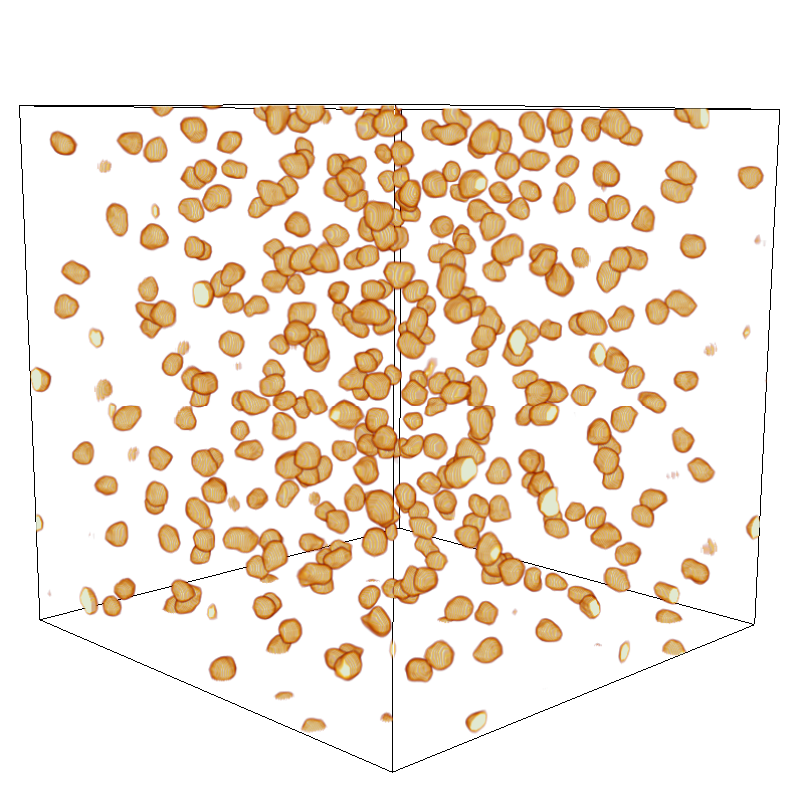} 
& \includegraphics[width=0.2\textwidth]{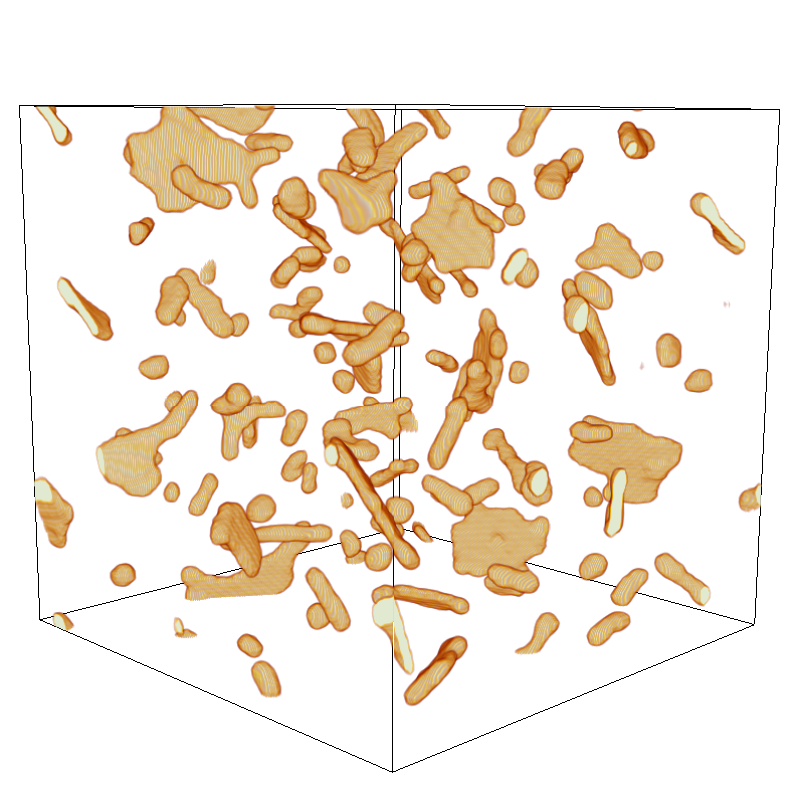}   
\end{tabular}
\end{ruledtabular}
\end{table*}

\subsection{Expansion Rate}\label{ssec:epsdot}

Due to the slow timescale of ejecta evolution in NS-mergers ($\sim 1 \unit{ms}$) relative to the timescale of nuclear reactions near saturation density, we study the role of the expansion rate on the configuration of our system at 1.0 MeV. Our goal was to identify the fastest expansion rate the simulations could experience before coming out of equilibrium. This allows us to minimize the computation times for the simulations in Sec. \ref{ssec:populations}.

We use our highest temperature for these simulations for three reasons. Firstly, it allows us compare our nuclide populations to available nuclear statistic equilibrium data obtained with a Virial expansion, which is more consistent with our classical approach. This is opposed to the Hartree-Fock calculations which have been done for lower temperatures. Secondly, it allows us to compare to our previous work \cite{PhysRevC.86.055805}. Lastly, the lower temperature data produced non-equilibrium structures, which are discussed in more detail in Sec. \ref{ssec:nonequilibrium}.

We simulate 51\,200 particles at a temperature of 1.0 MeV for expansion rates between $\dot\xi = 10^{-5} - 10^{-8}$ c/fm and evolve our simulations for $t = \dot\xi^{-1}$ starting at an initial density of $0.08 \unit{fm}^3$ down to $0.01 \unit{fm}^3$. The initial configurations are first equilibrated from random for $5 \times 10^5$ timesteps at a density of $n=0.08 \unit{fm}^{-3}$. Table \ref{Tab:expansionstats} compares the mass fraction of free neutrons, the mean mass number of clusters, and the mean charge number of clusters in our simulations to NSE data. The NSE data is obtained from the tables of G. Shen \etal \cite{PhysRevC.82.045802}, which include 8980 species of nuclei with $A\geq12$ as well as protons, neutrons and alpha particles, and shows fair agreement with the IUMD results.

Light isotopes ($A<12$) accounted for less than 1\% of the mass fraction of all simulations and are excluded from the analysis of mean cluster mass and charge number for comparison to NSE. The vast majority of these clusters were free protons with several bound neutrons.

We observe that the only simulations where clusters with $A>300$ were observed when proton fractions $Y_P=$0.3, 0.4 were stretched at $\dot\xi = 10^{-5}$ c/fm. These cases were not able to equilibrate due to the short time of the simulation (50\,000 timesteps). In the $Y_P$=0.3 case the mass fraction of the system above $A>300$ is 10\%, and there are many large oblong clusters that are approaching fission. In contrast, more than 66\% of the mass of the system in $Y_P$=0.4 case are in clusters where $A>300$, the largest being $A=3063$. This system is still in the spaghetti phase at $n=0.01 \unit{fm}^{-3}$.

We find that expansion rates of $\dot\xi = 10^{-5}$ c/fm are too fast and produce non-equilibrium effects, while expansion rates of $\dot\xi = 10^{-6}$ c/fm and $\dot\xi = 10^{-7}$ c/fm do not differ much from each other and are in good agreement with NSE data. For the one case we tested $\dot\xi = 10^{-8}$ c/fm, we also find good agreement. For this reason, we choose $\dot\xi = 10^{-7}$ c/fm as our expansion rate for the main simulations in \ref{ssec:populations} as they can be computed in a reasonable amount of time while still maintaining statistical equilibrium.

\subsection{Nuclide Abudances at 0.75 and 1.0 MeV}\label{ssec:populations}

\begin{figure*}[t]
\centering

\includegraphics[width=1.0\textwidth]{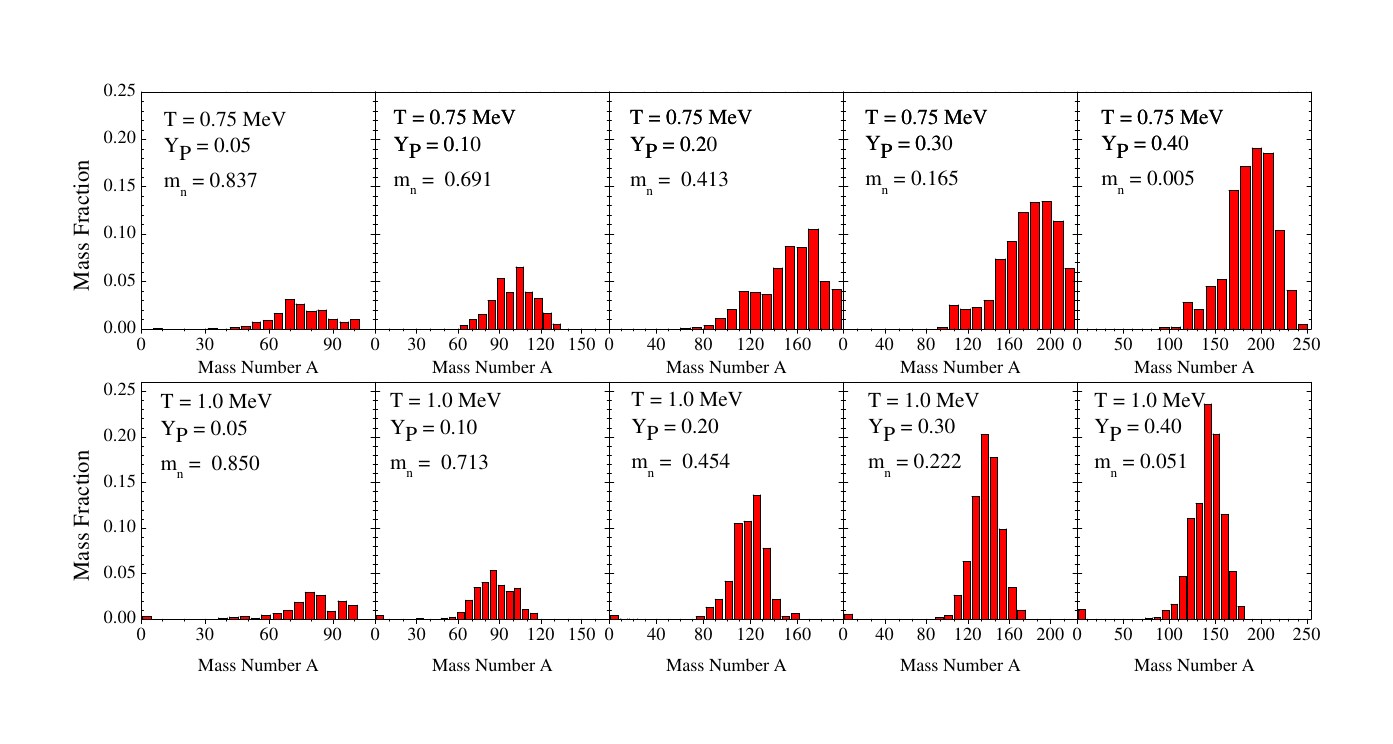}
\caption{\label{fig:MassHist075}  Distribution of cluster masses for $T=0.75 \unit{MeV}$ (top row) and $T=1.0 \unit{MeV}$ (bottom row) for proton fractions $Y_P =$ 0.05, 0.10, 0.20, 0.30, and 0.40 (columns) from simulations of $51\,200$ nucleons at a density of $n=0.00125 \unit{fm}^{-3}$. The mass fraction of free neutrons $m_n$ is indicated.}

\includegraphics[width=1.0\textwidth]{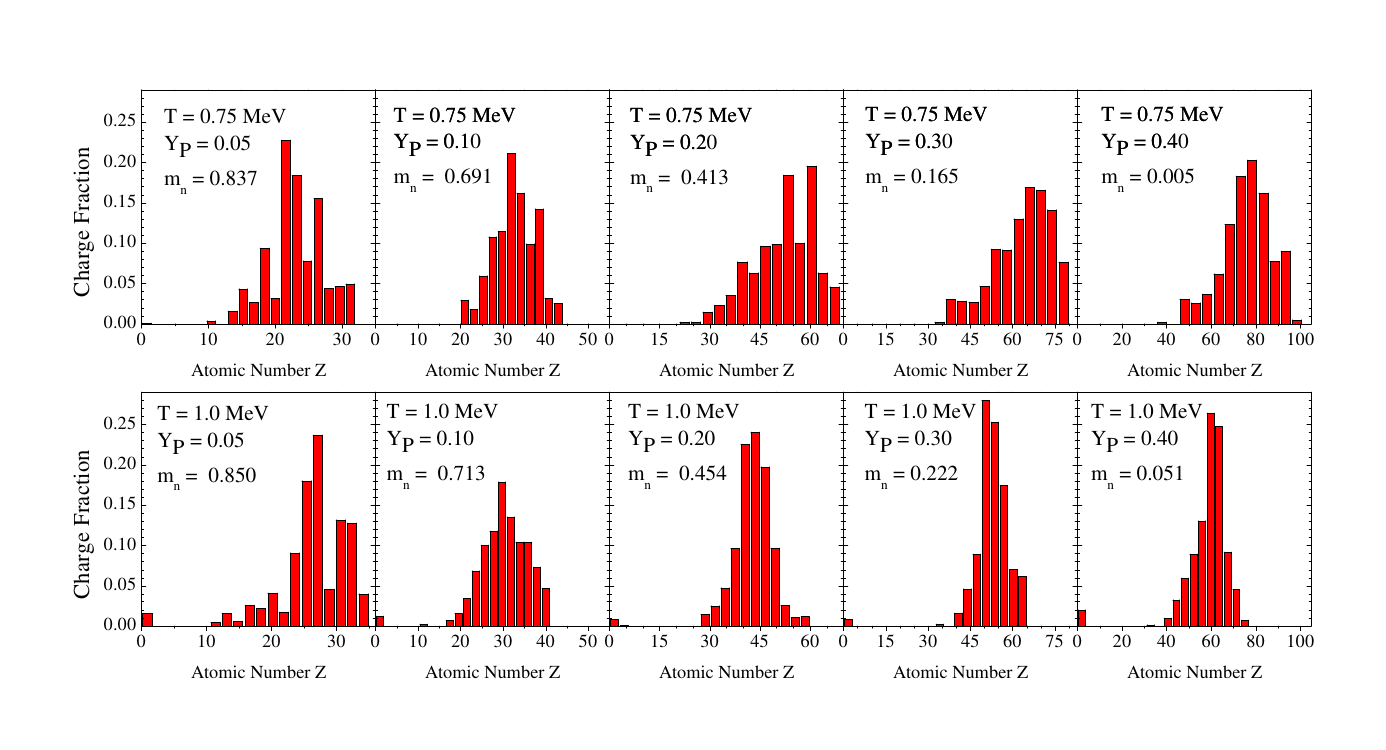}
\caption{\label{fig:CharHist075}  Distribution of charge for $T=0.75 \unit{MeV}$ (top row) and $T=1.0 \unit{MeV}$ (bottom row) for proton fractions $Y_P =$ 0.05, 0.10, 0.20, 0.30, and 0.40 (columns) from simulations of $51\,200$ nucleons at a density of $n=0.00125 \unit{fm}^{-3}$.  The mass fraction of free neutrons $m_n$ is indicated.}

\end{figure*}

\begin{figure*}[t]
\centering

\includegraphics[width=1.0\textwidth]{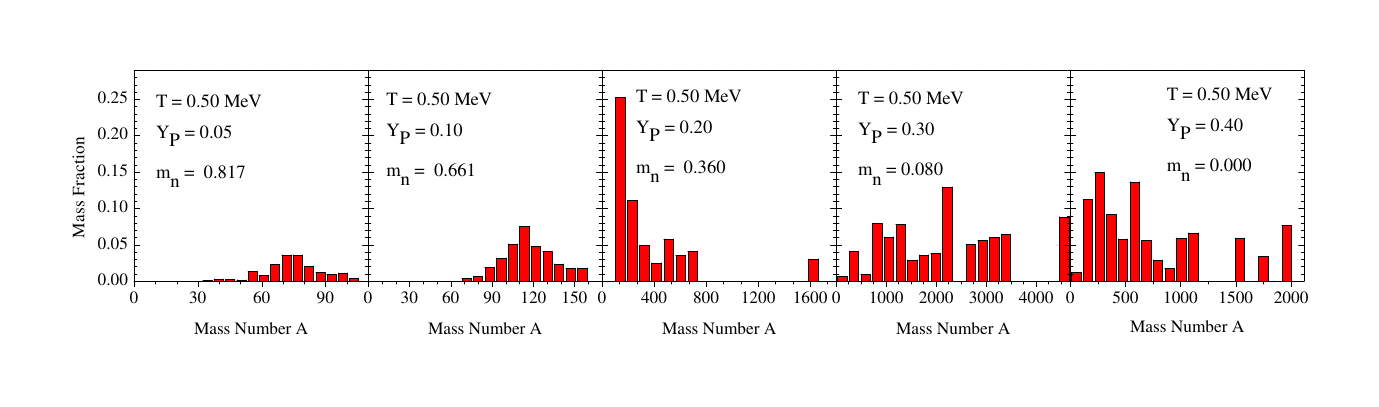}
\caption{\label{fig:MassHist05}  Distribution of cluster masses for $T=0.5 \unit{MeV}$ for proton fractions $Y_P =$ 0.05, 0.10, 0.20, 0.30, and 0.40 from simulations of $51\,200$ nucleons at a density of $n=0.00125 \unit{fm}^{-3}$. The mass fraction of free neutrons $m_n$ is indicated.}

\includegraphics[width=1.0\textwidth]{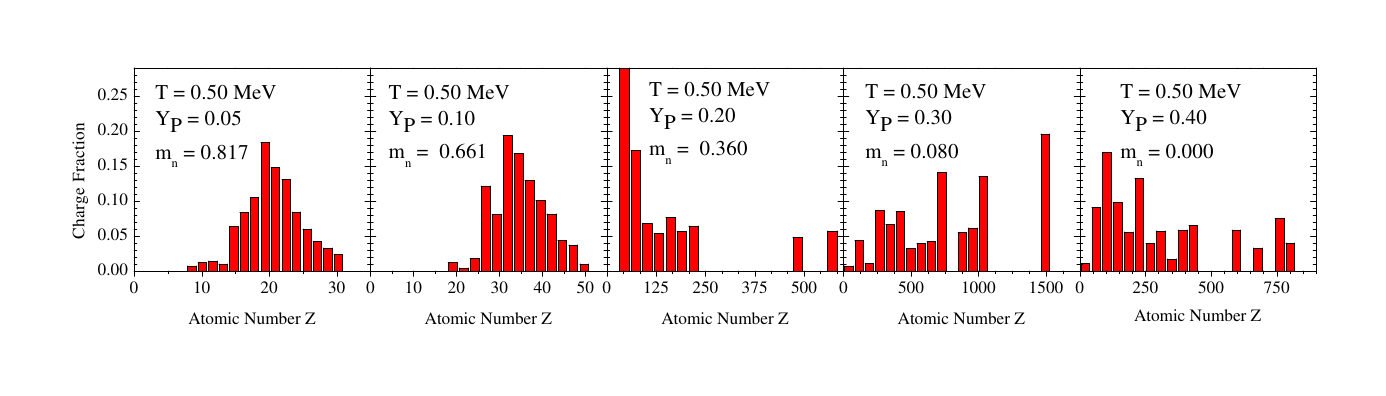}
\caption{\label{fig:CharHist05}  Distribution of charge for $T=0.5 \unit{MeV}$ for proton fractions $Y_P =$ 0.05, 0.10, 0.20, 0.30, and 0.40 from simulations of $51\,200$ nucleons at a density of $n=0.00125 \unit{fm}^{-3}$.}

\end{figure*}

Our simulations survey five proton fractions, $Y_{P} =$ 0.05, 0.10, 0.20, 0.30, and 0.40, and three temperatures, $T = $ 0.5, 0.75, and 1.0 MeV.  All 15 cases were simulated using 51\,200 particles, and the runs with $Y_{P} = 0.05$ were also run with 409\,600 particles to get better statistics for the mean number of protons in each cluster. These simulations are evolved for $3 \times 10^7 \unit{fm/c}$ with 2 fm/c per timestep on 32 GPU nodes on Big Red II, and are expanded at a rate of $\dot\xi_i = 10^{-7}$ c/fm, starting at an initial nucleon density of 0.08 fm$^{-3}$ decreasing down to 0.00125 fm$^{-3}$.

Our simulations are equilibrated from uniformly distributed random initial positions for $5 \times 10^5$ timesteps at a density of $n=0.08 \unit{fm}^{-3}$.  The higher proton fractions of 0.3 and 0.4 begin as lasagna, while the lower proton fractions below 0.2 have less regular structure and are highly dependent on the temperature. During the expansion, we observe that the pasta phase transitions occurred at densities consistent with the results of our previous work \cite{PhysRevC.88.065807}. Between densities of $0.02 \unit{fm}^{-3}$ and $0.01 \unit{fm}^{-3}$ the spaghetti fissions to produce the spheroidal gnocchi, whose statistics are presented in Fig. \ref{fig:MassHist075} and Fig. \ref{fig:CharHist075}, and is visualized in Tab. \ref{tab:Visualizations}. We specifically observe that the gnocchi clusters form a lattice in the $Y_P = 0.40$, $T=1.0$ MeV simulation near densities of $n=0.01 \unit{fm}^{-3}$, which is consistent with our previous work. 

Below densities of 0.01, the proton populations of individual gnocchi stayed relatively constant, with new clusters forming only occasionally when protons escape from clusters. This is more common at 1.0 MeV, which produced larger populations of light isotopes ($A < 12$) than at 0.75 MeV.

The free neutron population grew asymptotically in all cases, and eventually flattens out close to when our simulations end. We expect that our final cluster masses would continue to shed neutrons as the simulations evolved, but we expect a change of no more than 1\% in their mass fractions. 

In our final configurations we observe several trends. First, the distribution of cluster masses and charges is approximately Gaussian. Comparing the temperatures across proton fractions, the mean mass and charge of clusters was greater in the 0.75 MeV simulations; conversely, the mass fraction of free neutrons was greater in the the 1.0 MeV simulations. Similarly, comparing results at constant temperature reveals that both the mean number of protons per cluster and the mean mass of clusters grows with increasing proton fraction.

\begin{figure}[h]
\centering
\includegraphics[width=0.5\textwidth]{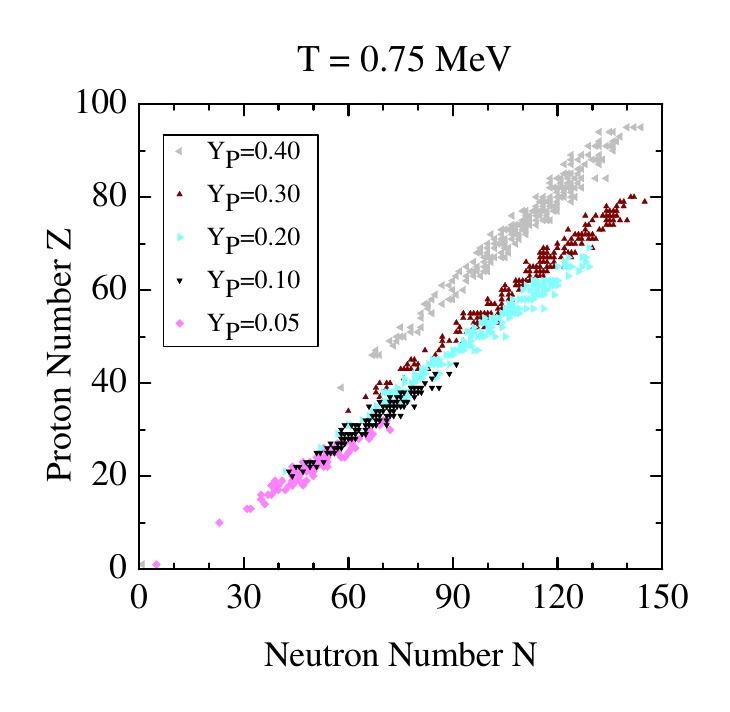}
\caption{\label{fig:Nuclides075} (Color online) Atomic number $Z$ and neutron number $N$ of nuclides for simulations at $T=$ 0.75 MeV with $Y_P = $ 0.05, 0.10, 0.20, 0.30 and 0.40 described in Sec. \ref{ssec:populations}.}
\end{figure}

\begin{figure}[h]
\centering
\includegraphics[width=0.5\textwidth]{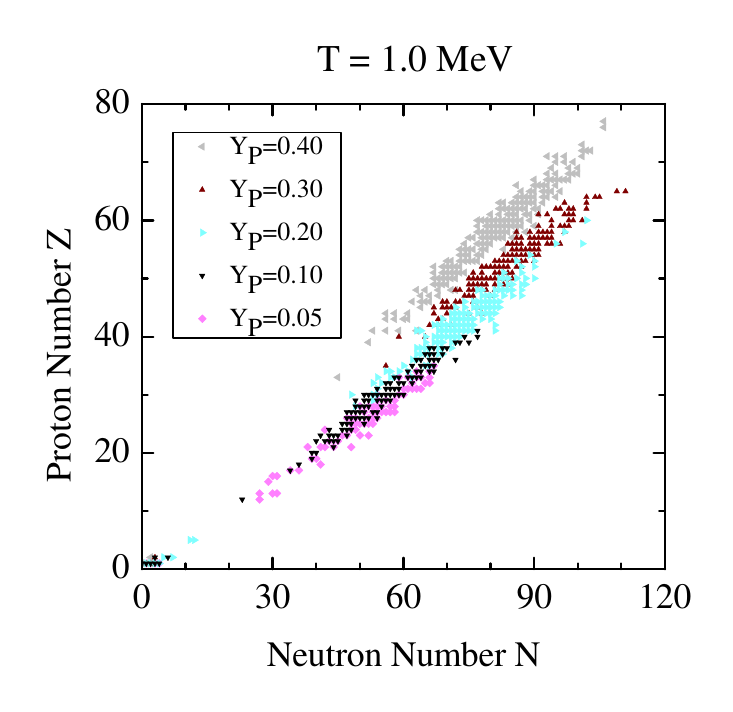}
\caption{\label{fig:Nuclides1} (Color online) Atomic number $Z$ and neutron number $N$ of nuclides for simulations at $T=$ 1.0 MeV with $Y_P = $ 0.05, 0.10, 0.20, 0.30 and 0.40 described in Sec. \ref{ssec:populations}.}
\end{figure}

We observe an approximate ``2 to 1 rule" for mass fractions of bound neutrons. At both temperatures, there are approximately two bound neutrons for every proton. For example, in a system with 5\% protons approximately 15\% of nucleons will be bound in clusters, with the remaining 85\% of the system as free neutrons. This lets us predict that systems with $Y_{P} > 0.33 $ will have a negligible population of free neutrons, as well as the slope of the N vs Z plots of nuclides present in our clusters, in Fig. \ref{fig:Nuclides075}.

Consistent with our observations in Sec. \ref{ssec:epsdot}, we also observe the presence of light isotopes at $T=$ 1.0 MeV which account for less than 1\% of the total mass, but we find no such clusters at $T=$ 0.75 MeV. These can be seen in Fig. \ref{fig:MassHist075} and Fig. \ref{fig:Nuclides1}.

\subsection{Nuclide Abudances at 0.5 MeV}\label{ssec:nonequilibrium}

Much of the discussion offered for the simulations at $T=$ 0.75 and 1.0 MeV apply to the simulations at $T=$ 0.5 MeV at proton fractions of $Y_P = $ 0.05 and 0.10. There is a slight increase in the mean mass and mean charge of clusters, with a decrease in the mass fraction of free neutrons when compared to the simulations at  $T=$ 0.75 and 1.0 MeV. This can be seen in Fig. \ref{fig:MassHist05} and Fig. \ref{fig:CharHist05}

The simulations at $T=$ 0.50 MeV with $Y_P \geq 0.20$ show non-equilibrium effects, with a static population of super heavy clusters. We find that this is the result of a phase transition in the nucleons. While the \textit{macrophase} of the system can be described by the pasta, the \textit{microphase} of the nucleons within the pasta structures is either liquid or solid. At temperatures above $T=0.5$ MeV the nucleons undergo a phase transition to the solid microphase while still at high densities, which produces a collection of large rods, slabs, and spheroids as seen in Tab. \ref{tab:Visualizations}. This was recently observed in molecular dynamics simulations of pasta by Alcain \etal \cite{Alcain:2014fma}, who found that the nucleons undergo a solid-liquid phase transition at low densities near $T=0.5$ MeV. Although our observation is consistent Alcain \etal, it is worth mentioning that the nuclear potential used there differs slightly from the one we use here. 

%
%
%

\subsection{Simulations with 409\,600 Nucleons}\label{ssec:409600}

\begin{figure}[h]
\centering
\includegraphics[width=0.5\textwidth]{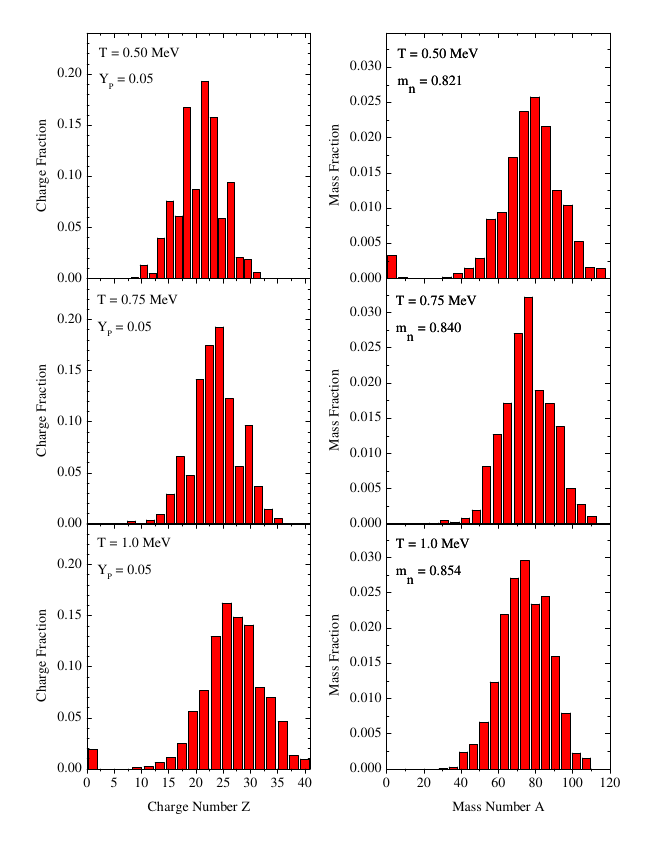}
\caption{\label{fig:409600} (Color online) Charge fractions (left) and mass fractions (right hand side) for simulations of 409\,600 particles at $T=$ 0.5, 0.75, and 1.0 MeV with $Y_P = $ 0.05 as described in Sec. \ref{ssec:409600}.  The mass fraction of free neutrons $m_n$ is indicated on the left hand panels.}
\end{figure}

We perform simulations containing 409\,600 particles for three reasons: (1) to test the scaling of our code performance, which was discussed in Sec. \ref{sssec:performance}; (2) to check for finite size effects; and (3) to specifically obtain better cluster statistics for the $Y_P = 0.05$ cases, see Fig. \ref{fig:409600}. With 51\,200 particles at $Y_P = 0.05$ the simulation only contains 2\,560 protons, which yields the poor Gaussians in the first column in Figs. \ref{fig:MassHist075}, \ref{fig:CharHist075}, \ref{fig:MassHist05}, \& \ref{fig:CharHist05}. 

Aside from the poor cluster statistics at low proton fractions, we observe negligible finite size effects.  Simulations with 409\,600 particles are in good agreement with simulations using 51\,200 particles.  In particular, we observe that the mass fraction of free neutrons, mean cluster charge, and mean cluster mass are in good agreement with results found in Secs. \ref{ssec:populations} \& \ref{ssec:nonequilibrium}. 
 
\section{Conclusions}\label{sec:Conclusions}

We have performed molecular dynamics (MD) simulations of the decompressing nuclear matter that may be ejected during neutron star mergers.  We slowly expand a simulation volume and then analyze the final configurations for the different kinds of nuclei (clusters) present.  We find that, as long as the expansion rate is not too high and the temperature is not too low, the system appears to remain in statistical equilibrium.   In general, at temperatures of 0.75 and 1.0 MeV, expanding MD simulations at rates of $\dot\xi = 10^{-7} \unit{c/fm}$ or slower produces distributions of nuclei similar to many nuclear statistical equilibrium (NSE) models.  Expansion at faster rates do not allow the clusters to remain in equilibrium and produces a greater spread in the masses and charges of the final clusters.   At $T= $ 0.5 MeV with $Y_P \geq 0.20$ the system comes out of NSE by undergoing a solid-liquid phase transition at the nucleon scale, producing large rod like clusters which do not fission.

This simple MD model can describe matter over a large range of densities where the system may be uniform nuclear matter, a variety of complex nuclear pasta phases, or a collection of more or less isolated nuclei.  At lower densities we reproduce many features of NSE models, such as the mass fraction of free neutrons and the mean mass and charge of heavy nuclei.  Therefore our model can describe matter at high densities, during the initial stages as it is ejected during neutron star mergers, through latter stages where nuclei and free neutrons provide the initial conditions for more conventional nuclear reaction network calculations of nucleosynthesis.  

In future work we will use our simulations to predict a variety of weak interaction rates for charged lepton and neutrino capture on the complex nuclear pasta phases.  This may allow better predictions of the initial proton fractions and temperatures of neutron rich matter ejected in NS mergers. 

\begin{acknowledgments}

We are grateful to David Reagan at the Advanced Visualization Laboratory - Indiana University for his help with ParaView.  
We would also like to thank Indiana University for access to the BigRed II supercomputer.  
This research was supported in part by DOE grants DE-FG02-87ER40365 (Indiana University) and DE-SC0008808 (NUCLEI SciDAC Collaboration).

\end{acknowledgments}


%

\end{document}